\documentclass[usenatbib]{mn2e}
\usepackage{graphicx}
\title{Verification of the diffusive shock acceleration in Mrk 501}
\author[Zheng et al.]{Y.G. Zheng$^{1,2,3,4}$\thanks{E-mail: ynzyg@ynu.edu.cn}, G.B. Long$^{5}$; C.Y. Yang$^{2,4}$; J.M. Bai$^{2,4}$\thanks{Corresponding author. E-mail:  baijinming@ynao.ac.cn};\\
       $^{1}$Department of Physics, Yunnan Normal University, Kunming 650092, China\\
       $^{2}$Yunnan Observatories, Chinese Academy of Sciences, Kunming 650011, China\\
       $^{3}$Shandong Provincial Key Laboratory of Optical Astronomy and Solar-Terrestrial Environment£¬Shandong University, Weihai, 264209\\
       $^{4}$Key Laboratory for the Structure and Evolution of Celestial Objects, Chinese Academy of Sciences\\
       $^{5}$Department of Physics and Astronomy, Sun Yat-Sen University, Zhuhai 519082, China}
\begin{document}
\date{Received date  / Accepted date}

\pagerange{\pageref{firstpage}--\pageref{lastpage}} \pubyear{2018}

\maketitle

\label{firstpage}

\begin{abstract}
The present work considers a plane shock front propagating along a cylindrical jet. Electrons experience the diffusive shock acceleration around the shock front, and subsequently drift away into the downstream flow in which they emit most of their energy. Assuming a proper boundary condition at the interface between the shock zone and the downstream zone, we solve the transport equation for the electrons in the downstream flow zone, where the combined effects of escape, synchrotron and IC cooling in the Thomson regime are taken into account. Using the electron spectrum obtained in this manner we calculate the multi-wavelength spectral energy distribution of Mrk 501 in the synchrotron self-Compton scenario. We check numerically if the Klein-Nishina cross-section could be approximated to the Thomsom regime. We consider whether the model results yield physically reasonable parameters, and further discuss some of implications of the model results. It suggests that the process of diffusive shock acceleration operates in the outflow of Mrk 501.
\end{abstract}


\section{Introduction}

Multi-wavelength observations show that the spectral energy distribution (SED) of blazars exhibits double-hump shape, with one hump extending in frequency from the radio to the ultraviolet, and in extreme cases to keV X-rays (Costamante et al. 2001) and the other covering the $\gamma$-ray energy regime. It is generally believed that the low-energy hump is produced by synchrotron radiation from relativistic electrons in the jet (Urry 1998). The mechanism producing the high-energy hump is an open issue. Probably it is produced by inverse Compton (IC) scattering of the relativistic electrons, either on the synchrotron photons (synchrotron self-Compton, SSC)(Maraschi et al. 1992) or on some other photon populations (External Compton, EC) (Dermer \& Schlickeiser 1993; Sikora et al. 1994). In the framework of blazar paradigm, the observed SED could be reproduced by the model, in which a non-thermal relativistic electron energy distribution (EED) is assumed (Dermer 2015). The EEDs used in various calculations include: 1) a simple power law (e.g. Katarzynski et al. 2006); 2) broken power law (e.g. Tavecchio et al. 1998; Katarzynski et al. 2001; Albert et al. 2007; Tavecchio et al. 2010; Zheng \& Kang 2013); 3) log-parabolic (e.g. Massaro et al. 2004; Tramacere et al. 2011; M. Hayashida et al. 2012; Chen et al. 2014); 4) power law with an exponential high-energy cutoff (e.g. Finke et al. 2008; Yan et al. 2013); 5) power law at low energies with a log-parabolic high-energy branch(e.g. Massaro et al. 2006; Tramacere et al. 2009); 6) double broken power law (e.g. Abdo et al. 2011a; 2011b).

These EEDs reconstructed from the observed emission can potentially be explained by the \emph{Fermi} type acceleration mechanisms (e.g. Paggi 2010; Tramacere et al. 2011), and they can be calculated assuming power-law injection from the shock and energy losses and escape (or adiabatic expansion) in the region downstream from the shock (e.g. Mastichiadis \& Kirk 1997; Kusunose et al. 2000).

The supersonic outflows in a blazar jet naturally generate shocks, and these could form the principal region where the kinetic energy of the jet flow is dissipated via acceleration of electrons to the relativistic energies required for emission of X-rays and $\gamma$-rays. Diffusive shock acceleration (DSA) is a potentially efficient mechanism for producing energetic particles from a flow with strong shocks (Kirk et al. 1998). It is believed to be an important acceleration mechanism in blazars (e.g. Bell 1978; Blandford \& Ostriker 1978; Drury 1983; Blandford \& Eichler 1987; Jones \& Ellison 1991). Such acceleration only produces a strict power-law energy spectrum of energetic particles in the test-particle approximation, where the back-reaction of the pressure of the accelerated particles on the flow is neglected. A more complicated spectrum is caused by competition between the acceleration (or and injection) and escape (or and cooling) from the shock region. In these scenarios, We could expect some distinctive properties on DSA in both EED and induced SED. Theoretical studies of DSA (e.g. Kirk \& Heavens 1989; Ellison et al. 1990; Ellison \& Double 2004a; Summerlin \& Baring 2012) suggest that a wide variety of power-law indices for the accelerated charged particles are possible for a given velocity compression ratio across the discontinuity. These indices are sensitive to the orientation of the mean magnetic field, and the character of the \emph{in situ} magnetohydrodynamic (MHD) waves (Baring et al 2017).

We note that determining the jet physics and environment from SED of blazars is a tricky problem of inversion (Dermer et al. 2014). Due to observational limitations, We can not obtain a detailed picture of the blazar region. Since accelerated non-thermal particles can produce the broadband continuum radiation that is detected from blazar jets, we could use the SED to diagnose a particle acceleration process in blazars. For instance: DSA is potentially efficient that only low levels of magnetic turbulence are required in blazar jets to accommodate synchrotron spectral peaks appearing in the X-ray band (Inoue \& Takahara 1996).

A standard model with a magnetized plasma ejected at relativistic speed in a collimated outflow along the polar axes of a rotating black hole has been developed to explain the observational features of blazars (B$\rm\ddot{o}$ttcher et al. 2012). Since BL Lac objects are characterized by an almost complete absence of emission lines or only weak emission lines (e.g. Urry \& Padovani 1995), in the lepton model framework, the SSC scenario is an exclusive explanation for the observed SEDs from high synchrotron peaked BL Lacs (HBLs). Using homogeneous SSC model spectral to the extensive multi-wavelength observations of typical HBLs (e.g. Bednarek \& Protheroe 1999; Katarzynski et al. 2001; Tavecchio et el. 2001), both the energy spectrum and variability can be explained (e.g. Finke et al. 2008; Ghisellini et al. 2009). However, even a comprehensive investigation using the model suffers in both characterizing the non-thermal EED and dominating the acceleration mechanism (Kataoka et al. 1999; Aharonian et al. 2009; Abdo et al. 2011a; 2011b).  Assuming two spatial zones, the present paper focus on shaping the non-thermal EED at blazar jet shocks. The aim of the paper is to determine whether the EED resulting from the DSA process can reproduce the multi-wavelength SED of Mrk 501 in the homogeneous SSC scenario. Throughout the paper, we assume the Hubble constant $H_{0}=75$ km s$^{-1}$ Mpc$^{-1}$, the dimensionless numbers for the energy density of matter $\Omega_{\rm M}=0.27$, the dimensionless numbers of radiation energy density $\Omega_{\rm r}=0$, and the dimensionless cosmological constant $\Omega_{\Lambda}=0.73$.

\section{Particle Spectrum}
We consider a plane shock front with a compression ratio $r=u_1/u_2$ propagating along a cylindrical jet of constant cross-section, where, $u_1$ is the speed of upstream flow and  $u_2$ is the speed of downstream flow. Electrons are accelerated at the shock front, and subsequently drift away from it into the downstream flow. This scenario can be treated as there are two spatial zones (e.g. Ball \& Kirk 1992): one around the shock front, in which particles are continuously accelerated, and the other downstream of shock, in which particles emit most of their energy. We assume that both zones contain relativistic electrons distributed isotopically in momentum space and a uniform magnetic field.

\subsection{The injection spectrum at the shock front}
The details of particle transport around the shock front are given in Dermer \& Menon (2009). In the case of a non-relativistic and parallel shock (magnetic field along the shock normal), we assume the particle distributions satisfy $N_{1}(\gamma)\propto\delta(\gamma-\gamma_{0})$ in the upstream region, where the $\gamma_{0}$ is characteristic energy of the particles. If the size of shocked flow is limitless, using the zero flux boundary condition, we could obtain a steady particle spectrum in the downstream flow (Dermer \& Menon 2009)
\begin{equation}
N_{2}(\gamma)=N_{0}\gamma^{-s}~~~~~{\gamma_1} \le \gamma \le{\gamma _2}\;,
\label{Eq:1}
\end{equation}
where, the spectral index $s=(r+2)/(r-1)$.

\subsection{The resulting EED in the downstream flow}
Assuming that accelerated particles $N_{2}(\gamma)$ subsequently drift away from shock front into the downstream flow, where they lose energy, we write $Q(\gamma)\sim\chi N_{2}(\gamma)$. Here, $\chi$ is a proper boundary condition at the interface between the shock zone and the downstream zone in the unit of $s^{-1}$. In this scenario, we can evolve the energetic particle distribution with Lorentz factor between $\gamma$ in the downstream flow (e.g. Kirk et al. 1998)
\begin{equation}
\frac{{\partial N(\gamma,t)}}{{\partial t}}=\frac{\partial}{{\partial\gamma }}\left( {\frac{d\gamma}{dt}}\right)N(\gamma,t)-\frac{N(\gamma,t)}{{t_{esc}}}+Q(\gamma)\;.
\label{Eq:2}
\end{equation}
Where, in order to simplify the equation,  $t_{esc}^{-1}$ is assumed as an energy independent rate of particles escaping from emission region (e.g. Zheng et al. 2014), $d\gamma/dt=(d\gamma/dt)_{\rm syn}+(d\gamma/dt)_{\rm IC}$ describes the synchrotron and inverse-Compton (IC) cooling of the particles at time $t$. $(d\gamma/dt)_{\rm syn}=[4\sigma_{\rm T}/(3m_{\rm e}c)]U_{\rm B}\gamma^2$ is the rate of the synchrotron loss with the energy densities of the  magnetic field $U_{\rm B}$, electron rest mass $m_{\rm e}$, light speed $c$ and Thomson cross section $\sigma_{\rm T}$. Klein-Nishina (KN) effects can modify the electron distribution (e.g., Moderski et al. 2005; Nakar et al. 2009), the rate of IC energy losses in which the KN corrections are included is given by (Moderski et al. 2005):
\begin{equation}
(\frac{d\gamma}{dt})_{\rm IC}=\frac{4\sigma_{T}}{3m_{e}c}U_{\rm rad}F_{\rm KN}(\gamma)\gamma^{2}\;,
\label{Eq:3}
\end{equation}
where, $U_{\rm rad}=\int_{\epsilon_{\rm 0,min}}^{\epsilon_{\rm 0,max}}u(\epsilon_{\rm 0})d\epsilon_{\rm 0}$ is the total energy density of the radiation field, $u(\epsilon_{\rm 0})$ is the energy distribution of the soft photons, $\epsilon_{\rm 0}$ is the energy of soft synchrotron photons in units of $m_{e}c^2$, $F_{\rm KN}(\gamma)=\int_{\epsilon_{\rm 0,min}}^{\epsilon_{\rm 0,max}}f_{\rm KN}(\kappa)u(\epsilon_{\rm 0})d\epsilon_{\rm 0}/U_{\rm rad}$, $\kappa=4\gamma\epsilon_{\rm 0}$, the function $f_{\rm KN}(\kappa)$ can be approximated as (Moderski et al. 2005) as follow:
\begin{equation}
f_{\rm KN}(\kappa)\simeq\left\{ \begin{array}{ll}
1          & ~\kappa\ll 1 ~\mbox{(Thomson limit)}\\
\frac{9}{2\kappa^2}(\ln\kappa-\frac{11}{6})     & ~ \kappa\gg 1~ \mbox{(KN limit)}\;.
\end{array} \right.
\label{Eq:4}
\end{equation}
When $\kappa\leq 10^{4}$, $f_{\rm KN}(\kappa)\simeq 1/(1+\kappa)^{3/2}$. Therefore, the radiative cooling parameter is given by
\begin{equation}
\frac{d\gamma}{dt}=\frac{4}{3}\frac{\sigma_{\rm T}}{m_{\rm e}c}[U_{\rm B}+U_{\rm
rad}F_{\rm KN}(\gamma)]\gamma^2\;.
\label{Eq:5}
\end{equation}

Both the analytic and numeric approaches were adopted to solve the evolution equation (e.g. Kirk et al. 1994; Kirk et al. 1998; Chiaberge \& Ghisellini 1999; B$\rm\ddot{o}$ttcher et al. 2002; Zheng \& Zhang 2011; Finke \& Becker 2014). The analytic approach provides important insights into global properties of possible solutions, though it is often restricted to treating particles well above thermal energies where the acceleration process no momentum scale (Baring et al. 2016). In order to comprehend the physical effect that determines the spectrum, we solve the analytic solution of the equation in this special case. In the steady state, when energetic particles injection is balanced by losses and escape, the solution is (Dermer \& Menon 2009)
\begin{eqnarray}
N\left(\gamma \right)=\frac{1}{C_{0}\gamma^2}\int_{\gamma}^{\gamma_2} d\gamma'Q(\gamma')\exp\left(- \int_\gamma ^{\gamma'}\frac{d\gamma''}{C_{0} t_{esc}\gamma''} \right)\;£¬
\label{Eq:£¶}
\end{eqnarray}
Here we define the constant $C_{0}\sim c(\gamma)=4\sigma_{\rm T}[U_{\rm B}+U_{\rm
rad}F_{\rm KN}(\gamma)]/(3m_{\rm e}c)$. We determine the critical energy $\gamma_{c}=1/(C_{0}t_{esc})$ which follows from the balance between escape and cooling with $t_{esc}=t_{loss}(\gamma)$.

\begin{figure}
	\centering
		\includegraphics[angle=0,width=10 cm]{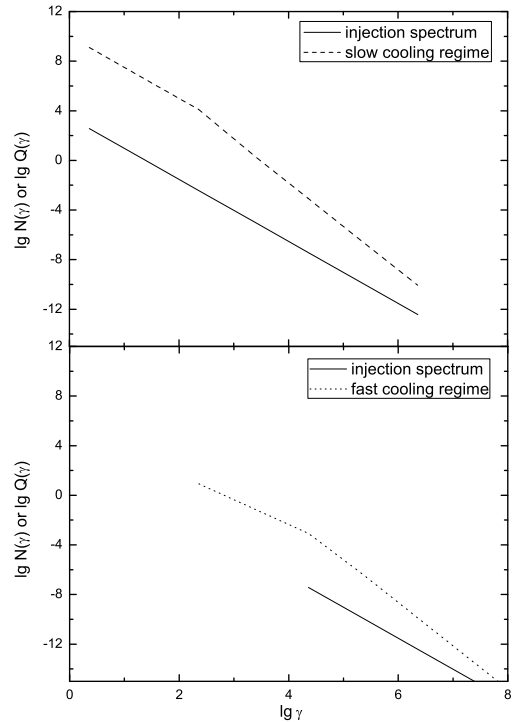}
		
	\caption{The injected spectrum and induced EED in both the slow cooling regime (top panel) and the fast cooling regime (bottom panel) as a function of the particle energy. We exhibit the injection spectrum as solid line. The dashed and dotted line shows the resulting EED in the slow cooling regime and the fast cooling regime, respectively. We adopt the parameters as follows: $ Q_{0}=3000\rm~cm^{-3}~s^{-1}$, $B=0.1\rm~G$, $F_{KN}\simeq1$, $s=2.5$, $t_{esc}=10^{17}/c\rm~s$, $U_{rad}=100U_{B}\rm~erg~cm^{-3}$.}
	\label{fig:1}
\end{figure}

We define the regime of fast and slow cooling using $\gamma_{c}$ and the minimum energy $\gamma_{1}$, with slow cooling if $\gamma_{1}<\gamma_{c}$ and fast cooling if $\gamma_{1}>\gamma_{c}$. We can approximate the EED in the slow cooling regime as
\begin{equation}
N\left( \gamma  \right) \approx \left\{ \begin{array}{l}
\chi N_{0}t_{esc}\gamma ^{-s}\;\;\;\;\;\;\;\gamma_{1}\le\gamma\le\gamma_{c}\\
\frac{\chi N_{0}}{C_{0}}\gamma ^{ - \left( {s + 1} \right)}\;\;\;\;\gamma_{c} < \gamma << \gamma_{2}
\end{array} \right.\;,
\label{Eq:£·}
\end{equation}
and in the fast cooling regime as
\begin{eqnarray}
N\left( {\gamma ,t} \right) \approx \left\{ \begin{array}{l}
\frac{\chi N_{0}\gamma_{1}^{(1-s)}}{C_{0}}\gamma ^{-2}\;\;\;\;\;\;\;\gamma_{c}\le\gamma\le\gamma_{1}\\
\frac{\chi N_{0}}{C_{0}}\gamma^{ -\left(s+1\right)}\;\;\;\;\gamma_{1}<\gamma << \gamma_{2}
\end{array} \right.\;.
\label{Eq:£¸}
\end{eqnarray}
In Figure {\ref{fig:1}}, we show the EED in the regime of fast- and slow cooling. It can be seen that: 1) in the slow cooling regime, the competition between the injection and the escape and/or cooling produces a power-law distribution that extends from the minimum energy up to maximum energy of the injection electrons. Below the critical energy, the energy losses are dominated by escape. This results in that the spectrum shape of electrons in this energy range is in consonance with that of injection spectrum. Above the critical energy, however, the energy losses are dominated by cooling process. The produced electron spectrum in this energy range becomes softer than the injection spectrum. 2) in the fast cooling regime, a power-law distribution that extends from the critical energy up to maximum energy of the injection electrons can also be produced, while we leave out of account the effects of escape. Below the minimum injection energy of the electrons, the index of the particle energy spectrum in this energy range is constant, and equal to 2. Above the minimum energy of the injection electrons, the cooling process also results to a softer electron spectrum than the injection spectrum.

\section{Modeling the SED of Mrk 501}
The BL Lac object source Mrk 501, at redshift $z=0.031$, is known as a typical HBL. Despite the fact that multi-wavelength SED of Mrk 501 have been extensively studied for quite some time (e.g., Bednarek \& Protheroe 1999; Katarzynski et a. 2001; Tavecchio et al. 2001; Kino et al. 2002; Albert et al. 2007; Acciari et al. 2011; Lefa et al. 2011; Zheng \& Zhang 2011; Mankuzhiyil et al. 2012; Neronov et al. 2012; Peng et al. 2014; Aleksic et al. 2015; Shukla et al. 2015; Aliu et al. 2016; Ahnen et al. 2017), the nature of these objects, such as the content of the jet, location and mechanism responsible for the $\gamma$-ray emission, are still far from being understood. Using the EED that is obtained in \S 2, we calculate the multi-wavelength SED of Mrk 501 in the homogeneous SSC scenario. We further discuss some of the implications of the model results. We interest to the characteristics of the EED can be used to constrain the microphysics expected for the DSA.

The homogeneous SSC radiation model that we adopt here assumes a spherical emitting blob with radius $R$ and comoving volume $V\simeq4\pi R^{3}/3$ in the downstream flow. It is filled by a uniform magnetic field, $B$, and extreme-relativistic electrons. We adopt a induced broken power-law function, Eq.(\ref{Eq:£·}), to describe the EED. Since Lewis et al. (2018) argues that the jet is not always in equipartition between the particles and magnetic field, we introduce a parameter $\eta_{\rm equi}$ to connect the non-thermal electron energy density $U_{e}$ and the magnetic field energy density $U_{B}$ in general(e.g. Zheng et al. 2017),
\begin{equation}
\eta_{\rm equi}=\frac{U_{e}}{U_{B}}=\frac{8\pi\int\gamma m_{e}c^{2}N(\gamma)d\gamma}{B^{2}}\;,
\label{Eq:9}
\end{equation}
Assuming isotropic distributions of synchrotron photons in the comoving frame, we evaluate the comoving synchrotron intensity $I_s(\nu)$ and the intensity of self-Compton radiation $I_{ic}(\nu)$, and then evaluate the flux density observed at the Earth as follows (e.g. Zheng \& Zhang 2011)
\begin{equation}
F_{\nu}=\pi\frac{R^{2}}{d_{L}^{2}}\delta^{3}(1+z)[I_s(\nu)+I_{ic}(\nu)]\;.
\label{Eq:10}
\end{equation}
Here, $d_{L}$ is the luminosity distance, and $\delta=[\Gamma(1-\beta\cos\theta)]^{-1}$ is the Doppler factor where $\Gamma$ is the blob Lorentz factor, $\theta$ is the angle of the blob vector velocity to the line of sight, and $\beta=v/c$.

Using the synchrotron and ICs solution for the spherical geometry, we can reproduce the SED of Mrk 501. In order to do so, we first determine the EED. Defining a density normalization coefficient $K=\chi N_{0}t_{esc}$, we rewrite the Eq. (\ref{Eq:£·}) as
\begin{equation}
N\left( \gamma  \right) \approx \left\{ \begin{array}{l}
K\gamma ^{-s}\;\;\;\;\;\;\;\gamma_{1}\le\gamma\le\gamma_{c}\\
K\gamma_{c}\gamma ^{ - \left( {s + 1} \right)}\;\;\;\;\gamma_{c} < \gamma << \gamma_{2}
\end{array} \right.\;.
\label{Eq:11}
\end{equation}
From Eq. (\ref{Eq:9}), we obtain a relation $K=4.85\times10^{4}\eta_{\rm equi}B^{2}[\gamma_{1}^{-(s-2)}/(s-2)-\gamma_{c}^{-(s-2)}/(s^2-3s+2)-\gamma_{c}\gamma_{2}^{-(s-1)}/(s-1)]^{-1}~cm^{-3}$. Then we calculate the EED between $\gamma_{1}=8.5\times10^{2}$ and $\gamma_{2}=6.0\times10^{6}$ with a break at critical energy $\gamma_{c}=5.6\times10^{5}$. Since the slope of the electron distribution at the shock is thought to be around 2 from the theory of the shock acceleration (Bell 1978; Bell et al. 2011; Summerlin \& Baring 2012), we set the energy index of the injection particles is $s=2.5$. The equipartition fraction $\eta_{\rm equi}$ depends predominantly on the minimum Lorentz factor of the radiating electrons. Hence, it is determined as $\eta_{\rm equi}\simeq178$ with the submillimeter flux included in the fitted data set.

We assume that relativistic electrons are in a steady state at the observational epoch. Therefore, we can calculate the SED within the framework of a homogeneous model using the determined EED. In order to do that, we determine the geometry and physical parameters of the emission region. It is well known that the comoving radius of the emission region is defined as $R\sim c\delta t_{var}/(1+z)$. Since the model focusses on the average behavior of Mrk 501, Abdo et al. (2011a) constrained the typical variability timescale $t_{var}\sim1-5$ days during the observational epoch. If we adopt the medial value of variability timescale $t_{var}\sim3$ days, we obtain $R=1.13\times10^{17}~cm$ for a Doppler factor $\delta=15$. On the other hand, radio observations of both the partially resolved core and submilliarcsec jet suggested that the magnetic field strength of Mrk 501 is $B=0.01-0.03~G$ (Giroletti et al. 2004; 2008). In the model, we adopt the low limit of magnetic field strength $B=0.01~G$.

\begin{figure}
	\centering
		\includegraphics[angle=0,width=9cm]{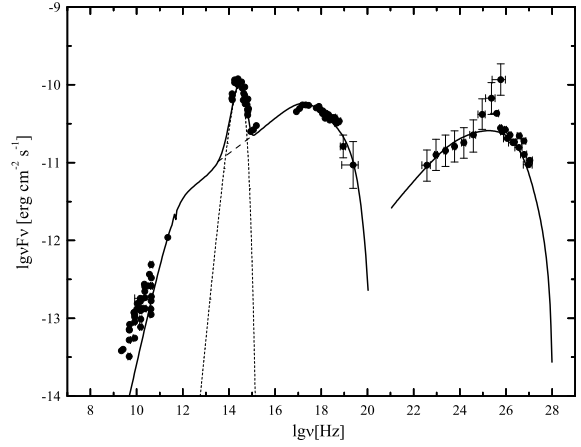}
		
	\caption{Comparisons of predicted multi-wavelength spectra with observed data of Mrk 501, averaged over all observations taken during the multi-wavelength campaign between 2009 March 15 and 2009 August 1. The quasi-simultaneous data are shown in solid circle. The dashed line represents the synchrotron emission, the dotted line represents inverse Compton emission on the synchrotron photons, the short dashed line estimates the starlight emission of the host galaxy using a given template in Silva et al. (1998), and the thick solid line represents total spectrum.}
	\label{fig:2}
\end{figure}

We show the predicted spectrum from radio frequencies to TeV $\gamma$ rays in Figure {\ref{fig:2}}. For comparison, the SED for Mrk 501 (Abdo et al. 2011a) averaged over all observations taken during the multi-wavelength campaign between 2009 March 15 and 2009 August 1 are also shown. In the figure, the quasi-simultaneous data are shown in solid circle. The dashed line represents the synchrotron emission, the dotted line represents inverse Compton emission on the synchrotron photons, the short dashed line estimates the starlight emission of the host galaxy using a given template in Silva et al. (1998), and the thick solid line represents total spectrum. The model parameters are listed in table {\ref{tb:1}}. It can be seen that the observed data can be reproduced in the model.

\begin{table*}
\begin{minipage}[t][]{\textwidth}
\caption{The physical parameters of the model spectra}
\label{tb:1}
\begin{tabular}{ll}
\hline\hline
parameters & broken power law EED\\
\hline\\
Magnetic field                                 & $B=0.01$~G                     \\
Emission region size                           & $R=1.13\times10^{17}$~cm       \\
Doppler factor                                 & $\delta=15$                   \\
Equipartition parameter                        & $\eta_{\rm equi}=U_{e}/U_{B}=178$   \\
Minimum electron energy                        & $\gamma_{min}=\gamma_{1}=850$ \\
Cooling electron break energy                  & $\gamma_{br}=\gamma_{c}=5.6\times10^{5}$   \\
Maximum electron energy                        & $\gamma_{max}=\gamma_{2}=6.0\times10^{6}$ \\
Index of injected electron                     & $s=2.5$ \\
Main variability timescale                     & $t_{var}\approx3$~days \\
Luminosity of the host galaxy                  & $L_{star}\approx3\times10^{44}\rm~erg~s^{-1}$ \\
Temperature of the host galaxy                 & $T_{star}=3500$~K  \\
\hline\\
\end{tabular}
\end{minipage}
\end{table*}

\section{Notes on the model results}
\begin{figure}
	\centering
		\includegraphics[width=9 cm]{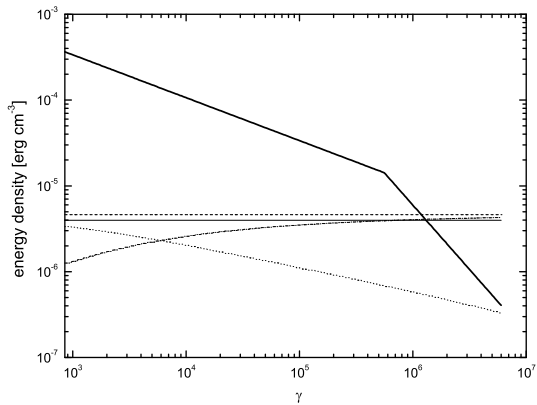}
		
	\caption{The predicted comoving energy density of ultra-relativistic electrons (thick solid curve). For comparison, the comoving energy density of the magnetic field (solid line), synchrotron photons (dashed line), synchrotron photons which are inverse Compton upscattered in the Thomson regime (dotted line), and synchrotron photons which are inverse Compton upscattered in the KN regime (dash dotted curve) are shown. }
	\label{fig:3}
\end{figure}

We interest to consider whether the model results yield physically reasonable parameters. In order to do that, we evaluated the predicted comoving energy density of ultra-relativistic electrons,
\begin{equation}
\gamma u_{e}(\gamma)=\gamma^{2}m_{e}c^{2}N(\gamma)\, .
\end{equation}
We show the comoving energy density of electrons for a given electron Lorentz factor in Figure {\ref{fig:3}}. For comparison, the comoving energy density of the magnetic field, $u_{B}\simeq3.9\times10^{-6}\rm erg~cm^{-3}$, synchrotron photons with a synchrotron emission coefficient $j_{\nu, syn}(\nu)$,
\begin{equation}
u_{syn}=\frac{4\pi R}{3c}\int j_{\nu, syn}(\nu)d\nu\simeq4.6\times10^{-6}\rm erg~cm^{-3}\,,
\end{equation}
synchrotron photons which are inverse Compton upscattered in the Thomson regime,
\begin{equation}
u_{syn, T}(\gamma)=\frac{4\pi R}{3c}\int^{m_{e}c^{2}/(4\gamma h)}j_{\nu, syn}(\nu)d\nu\,,
\end{equation}
and synchrotron photons which are inverse Compton upscattered in the KN regime,
\begin{equation}
u_{syn, KN}(\gamma)=u_{syn}-u_{syn, T}(\gamma)\,,
\end{equation}
are plotted in the figure as well. It can be seen that, 1) most of energy is stored in the low energy electrons with the mean Lorentz factor of the electrons $\gamma_{mean}=2447$; 2) even though the total energy density of the synchrotron photons is more than the energy density of the magnetic field, the dominant radiative cooling for all the electrons is attribution to synchrotron emission, Since $u_{syn/T}<u_{B}$ in all of the energy ranges; 3) Significantly, most of the synchrotron photons responsible for the gamma-ray emission are IC up-scattered in the KN regime. However, due to the low value of $F_{KN}(\gamma)$, IC scattering in the Thomson regime dominates the energy loss.

We now estimate the radiative efficiency of jet. Our model result suggests the shocked plasma is dominated by electron-proton pair within the emission region (Please see the next section for a further discussion). We adopt one electron-proton pair per electron-positron pair within the emission region (e.g. Celotti \& Ghisellini 2008). If we assume the mean Lorentz factor of the protons with $\gamma_{p, mean}\sim 1$, we find that the implied total jet power $L_{j}=L_{e}+L_{B}+L_{p}=\pi R^{2}c\delta^{2}(u_{e}+u_{B}+0.25u_{e})\sim2.46\times10^{44}\rm erg~s^{-1}$. We note that the total emitted radiative power is $L_{em}\simeq4\pi R^{2}c\delta^{2}(u_{syn}+u_{ssc})\sim7.35\times10^{42}\rm erg~s^{-1}$. This gives that the jet/blazar radiative efficiency was at the level of a few percent with $L_{em}/L_{j}=0.03$ in the observed epoches. Such a value for the radiative efficiency is commonly found in the jets of BL Lacs (e.g. Celotti \& Ghisellini 2008; Sikora et al. 2009).

On the other hand, reminding the relationship between the escape timescales and the critical energy of electrons, we could found $t_{esc}=1/(C_{0}\gamma_{c})$. If we adopt a typical escape timescales $t_{esc}=1.5R/c$ (e.g. Weidinger \& Spanier 2010), and adopt $U_{B}+U_{rad}F_{KN}(\gamma)\sim10^{-5}$, we estimate the radius of spherical emitting blob as $R\sim1.1\times10^{17}~\rm cm$ with the critical Lorentz factor $\gamma_{c}=5.6\times10^{5}$. This agrees with radius of model expected.

\section{Discussion}
In order to gain insight into the nature of the $\gamma$-ray emission, Abdo et al. (2011a) use a phenomenological model that introduces an EED with two different break energies for fitting the multi-wavelength SED of Mrk 501 during in a quiescent state. We note that this phenomenological EED can be interpreted in the following manner. One part with an electron energy less than the first break energy can be interpreted by DSA (Kirk 2000; Achterberg et al. 2001; Ellison \& Double 2004b; Keshet \& Waxman 2005; Sironi et al. 2015), and the other part is consistent with the EED where it is formed as the result of DSA, followed by cooling and escape (Blasi 2010; Zirakashvili \& Aharonian 2007; Vannonid et al. 2009), However, the physical origin of the first break energy is still not understand (Kakuwa et al. 2015). Since the model presented in the context assumes the injection spectrum due to DSA is transmitted without change to the downstream zone where the particles emit most of their energy, we can expect the resulting spectrum, after the effects of losses and escape have been calculated, contains clues about the injection spectrum produced by DSA, especially at low energies where the losses are small.

We adopt a proper boundary condition at the interface between the shock zone and the downstream zone to simplify the injection spectrum. The results of the SSC modeling presented in the previous section indicate that the energy spectrum of injection electrons is of the form $Q(\gamma)\propto\gamma^{-2.5}$ between electron energy $E_{\rm e, min}=\gamma_{1}m_{e}c^{2}\sim0.4~\rm GeV$ and $E_{\rm e, max}=\gamma_{2}m_{e}c^{2}\sim3100~\rm GeV$. We note that the formation of strong shock can be expected in the jet of Mrk 501 around the locations of a few parsecs from the core (Edwards \& Piner 2002; Piner et al. 2009). This distance scale could possibly be reconciled with the expected distance of the blazar emission zone from the center for the model parameters determined, $r=\Gamma R\sim0.4~\rm pc$ with a typical $\Gamma\sim10$ (e.g. Potter \& Cotter 2012; Zheng et al. 2017). In this scenario, we can expect for a DSA process at working within emission zone.

Generally, there are some electron populations with lower energies within the downstream emission zone, although their energy distribution has to be very flat or even inverted. In order not to overproduce the synchrotron radio photons and to response for the observed MeV-GeV $\gamma$-ray continuum. In these scenarios, predicted minimum electron energy should mark the injection threshold for the main acceleration mechanism, that is, only electrons with energies larger than $E_{e, min}$ are picked up by this mechanism to form the higher energy spectrum. On the other hand, the energy dissipation mechanisms operating at the shock fronts results to a injection energy scale (e.g. Peacock 1981). At this energy scale, particles are sufficiently energetic that they have a finite statistical probably of returning to the shock from the downstream region. Below this energy scale, even if the particles were beamed directly upstream along the magnetic field, they would still be traveling upstream in the local fluid frame slower than the downstream flow speed and would still, in the shock frame, be moving away from the shock. This makes it effectively impossible for particles with energies lower than this to get accelerated at the shock, because they never return to it once they have crossed it. It is believed that this injection energy scale in particular depends both on the thickness of the shock front and on the constituents of shock plasma (Abdo et al. 2011a). The shock thickness, in turn, is determined by the operating inertial length, or the diffusive mean free path of the particles, or both. Since this work can not determine the shock thickness, we limit the discussion to the constituents of the shocked plasma. We find that, if there are pure electron-positron pair plasmas, only the dynamic energy of a particle is contained, the threshold energy is set as $E_{\rm e, min}\sim\Gamma m_{e}c^{2}$. In contrast, if there are electron-proton pair plasmas, the shock thickness can be relatively enlarged, the DSA process operating electrons is demanded on higher energy with $E_{\rm e, min}\sim\epsilon\Gamma m_{p}c^{2}$. Here, $m_{p}$ is the mass of proton, and $\epsilon$ is the efficiency of the equilibration in the shock layer between shock thermal protons and their electrons counterparts (e.g. Abdo et al. 2011a), possibly, resulting from electrostatic potentials induced by charge separation of species of different masses (Baring \& Summerlin 2007). These give a model diagnostic for composition of the shocked plasma. A large minimum electron energy suggests that the shock plasma is dominated by electron-proton pair within the emission region. Using the minimum energy of shock-accelerated particles to constrain shock physics is interesting, but it is very model-dependent. Given the complexity of what the scattering waves, such as, Alfven waves, whistler waves, and so on, are responsible for confining the electrons near the shock, the issue should remain open.

The maximum attainable energy is determined by which of two conditions is met first. Either the shock runs out of time to accelerate particles or it runs out of space (e.g. Jokipii 1982; 1987). If one limits the lifetime of the shock, the rate of acceleration is an important, but not solely determining factor. The upper energy cut-off will be approximately the acceleration rate by the duration of the shock. If one limits the size of the shock, the upper energy cut-off will appear where diffusion length of particles is less than the scale size of the shock. Here diffusion length is a factor $c/v_{sh}$ larger than mean free path for scattering with the velocity of shock $v_{sh}$. In all cases, both of these are limiting factors, but usually one is the dominant limiter on the acceleration process. We note that the model assumes a particle drift paradigm, we can estimate the maximum energy of the electrons by equating the escape timescale $\tau_{esc}(\gamma)$ and the acceleration timescale $\tau_{acc}(\gamma)$, while an energy independent drift away rate is adopted. Assuming a mean free path $\ell(\gamma)$ for scattering with magnetic disturbances in a non-relativistic shock, the acceleration timescale could be derived in the test particle approximation, $\tau_{acc}(\gamma)=20\ell(\gamma)c/(3v_{sh}^{2})$ (e.g. Inoue \& Takahara 1996; Protheroe \& Clay 2004). The strongest possible MHD wave acceleration occurs in the \emph{Bohm limit} case when the $\ell(\gamma)$ is comparable to the Larmor radius $r_{L}=\gamma m_{e}c^{2}/(eB)$ (Krall \& Trivelpiece 1986). We neglect the advective escape process and assume that the escape occurs via spatial diffusion, $\tau_{esc}(\gamma)=r_{sh}^{2}/c\ell(\gamma)$, here $r_{sh}$ is radius of shock (e.g. Dermer \& Menon 2009; Kroon et al. 2017). Then, the maximum attainable energy in DSA is $E_{max}=1.16\times10^{8}(v_{sh}/c)(r_{sh}/10^{17})\rm GeV$. If one assumes that the shock front overruns the whole blazar region, we expect a shock with the velocity $v_{sh}\sim2.4\times10^{-5}c$.

The DSA expects a index of the electron spectrum to a value $s=(r+2)/(r-1)$. Our model as presented here predicts a softer spectrum with index $s=2.5$ resulting from a weakening compression ratio $r=3$, rather than the expected index of a strong shock front, $s=2$, in a gas whose ratio of specific heats is 5/3 (Drury 1983; Blandford \& Eichler 1987). We do not propose an explanation of why the shock has weakened. However, it is interesting to speculate that this might be the result of the back reaction of plasma which the shock has been accelerating. We could expect a softer spectrum with $s=2.5$ when the incorporation of anomalous transport properties associate with the wandering of magnetic field lines may efficiently reduce cross-field propagation (Kirk et al. 1996). On the other hand, relativistic shocks can produce a variety of power-law spectra including those with $s=2.5$. In relativistic shocks, the spectral index is dependent on multiple factors including the energy dependence of the mean free path, the nature of pitch angle scattering (large angle vs. small angle), the composition of the plasma (electron-positron or electron-proton), the equation of state, and even the magnetic field strength and obliquity.
\section{Conclusion}
Using the energy distribution of shock-accelerated electrons that are injected into the downstream region, assuming slow cooling, we are able to reproduce the multi-wavelength SED of Mrk 501 in a homogeneous SSC model. Our model indicated that: 1) A non-relativistic parallel shock with a compression ratio $r=3$ and with a velocity $v_{sh}\sim2.4\times10^{-5}c$ in the blazar region can produce a photon spectrum consistent with the observations. 2) The shock plasma in the emission region is mostly a hydrogen plasma. 3) Electrons are accelerated to a power-law energy distribution by diffusive shock acceleration and subsequently drift away from the shock into the downstream flow, where most of their energy is emitted. 4) The spectrum produced is consistent with synchrotron and IC cooling in the Thomson regime.

\section*{Acknowledgments}
We thank the anonymous referee for valuable comments and suggestions. This work is partially supported by the National Natural Science Foundation of China under grants 11133006, 11433004, 11463007, 11573060, the Key Research Program of the CAS (Grant NO. KJZD-EW-M06), and the Natural Science Foundation of Yunnan Province under grant 2016FB003, 2017FD072. Project Supported by the Specialized Research Fund for Shandong Provincial Key Laboratory. This work is also supported by the Key Laboratory of Particle Astrophysics of Yunnan Province (grant 2015DG035).


\end{document}